\documentclass[journal]{IEEEtran}
\usepackage{amsmath}
\usepackage{graphicx}
\usepackage{multirow}
\usepackage{nomencl}
\usepackage{cite}
\usepackage{graphicx,dblfloatfix}
\usepackage{epstopdf} 
\usepackage{algorithm}
\usepackage{algorithmic}
\usepackage{balance}
\IEEEoverridecommandlockouts
\begin{document}
	\renewcommand*\footnoterule{} 
	\title{Control Allocation for Wide Area Coordinated Damping}
	\author{M. \!Ehsan Raoufat,~\IEEEmembership{Student Member,~IEEE,} \!\!
		Kevin Tomsovic,~\IEEEmembership{Fellow,~IEEE,} and Seddik M. \!\! Djouadi,~\IEEEmembership{Member,~IEEE}
		\thanks{This work was supported in part by the National Science Foundation under grant No CNS-1239366, and in part by the Engineering Research Center Program of the National Science Foundation and the Department of Energy under NSF Award Number EEC-1041877 and the CURENT Industry Partnership Program.}
		\thanks{M. Ehsan Raoufat, Kevin Tomsovic and Seddik M. Djouadi are with the Min H. Kao Department of Electrical Engineering and Computer Science, The University of Tennessee, Knoxville, TN 37996 USA (e-mail: mraoufat@utk.edu).}}
	\maketitle
\begin{abstract}
\boldmath 
In this work, a modal-based sparse control allocation (CA) is proposed for coordinated and fault-tolerant wide-area damping controllers (WADCs). In our proposed method, the supervisory CA only communicates with necessary actuators to achieve the required damping performance and in case of actuator failures (e.g., due to loss of communication or scheduling), capabilities of the remaining actuators are fully used before the nominal performance is degraded. This method offers the advantages of modular design where WADC is initially designed to achieve satisfactory damping without the detailed knowledge of actuators. In the next step, CA is designed to manage actuator failures and limitations without the need to redesign the nominal WADC. The proposed approach is applied to a modified $286$-bus Western Electricity Coordinating Council (WECC) system to verify the feasibility on a complex power system. Simulation results indicate the effectiveness of the proposed method in coordinating multiple actuators and building resiliency.
\end{abstract}
\renewcommand\IEEEkeywordsname{Index Terms}
\section{Introduction}
\IEEEPARstart{I}{nter-area} oscillations have been identified as a major problem faced by most power systems and stability of these oscillations are of vital concern due to the potential for equipment damage and resulting restrictions on available transmission capacity between different areas \cite{PKundur}. With recent advances in wide-area measurement systems (WAMS), inter-area modes can be observed globally and wide-area damping controllers (WADCs) can be deployed to enhance the stability \cite{IKamwa}. Multiple design techniques and methodologies have been reported for damping inter-area oscillations including designs for supplementary control of generator excitation \cite{YZhang , MERaoufat_1, MMahmoudi}, FACTS devices \cite{SZhang_1,SZhang_2} and renewable energy sources \cite{ AELeon_1, AELeon_2}. However, very few such systems have been deployed in practice partly due to high level of robustness and reliability requirements for any closed loop power system controls. \\ \indent
Traditional power system topology is changing and a large number of small-scale renewable sources are being installed throughout the system. In this aspect, spatial distributions of wind farms are crucial to reduce the need for new transmission infrastructure. These wind farms could be selected as WADC actuators and contribute to damping inter-area oscillations though active$/$reactive power modulation \cite{AELeon_1}. In contrast with a large wind farm in a concentrated location, deployment of multiple small-scale wind farms will require special techniques for actuator coordination as none could be used individually to achieve adequate damping. Moreover, the availability of these weather dependent renewable resources could pose design challenges for reliability of critical controllers. \\ \indent
Considering reactive power modulation in Type 4 wind turbines (i.e., full converter asynchronous generators), the amount of available reactive power depends upon the operating mode, converter rating and grid code requirements. This may mean that some WADC actuators become temporarily unavailable (failed) or have more limited capabilities. Moreover, communication failures such as packet loss, excessive time delay and cyber-attacks may also lead to failures in these geographically-dispersed actuators. Thus, developing robust controllers to accommodate such failures and maintain the system stability is an important challenge in deploying WADCs. \\ \indent
In this paper, a sparse control allocation (CA) method is developed to optimally coordinate a set of actuators to damp the inter-area modes and achieve a fault-tolerant WADC. In our approach, the damping controller is designed based on a fault-free model and the supervisory CA distributes the control signals to necessary actuators based on the desired control actions, total cost, effects on different modes of the system and actuator constraints. This paper generalizes the previous methods on control allocation \cite{OHärkegard_1,OHärkegard_2,JBDavidson, MERaoufat_2} by considering the temporal sparsity and the effects of virtual control on the modal system. This technique allows us to give the highest priority to the control efforts associated with the critical inter-area modes. In \cite{ AELeon_1, AELeon_2}, an attempt to coordinate multiple wind farms was addressed but without considering the effects of actuator failures, capabilities and limits. This paper also extends \cite{IKamwa, YZhang, MMahmoudi, SZhang_1, SZhang_2} in which unavailability of WADC actuators has not been considered.  Feasibility of the proposed approach has been verified on a modified $286$-bus Western Electricity Coordinating Council (WECC) system with multiple small-scale wind farms.\\ \indent
This paper is organized as follows. In Section II, a modular control allocation technique is developed for system with redundant actuators and a multi-objective synthesis is presented as one method to design damping controller. Preliminaries on dynamic modeling of a WECC system with distributed wind farms are described in Section IV. Nonlinear time-domain simulations are presented in Section V to demonstrate the effectiveness of the proposed method in coordinating multiple actuators. Concluding remarks are given in Section VI.
\section{Modal-based Sparse Control Allocation for Wide-area Damping}
Control allocation can be used to coordinate a redundant set of actuators for a class of over-actuated systems in which the number of actuators ($m$) exceeds the number of states ($n$) \cite{OHärkegard_1}. Here, we consider model-based redundancy in the actuators as physical redundancy (e.g., replicating an actuator) is not cost effective in power systems. The assumption of redundancy $rank (B)=n<m$ needs to be satisfied to guarantee a set of admissible control signals \cite{OHärkegard_2}, where $B \in R^{n \times m}$ is the control input matrix. However, for power systems like many other practical systems, this assumption is not necessarily valid for the full-order system. In this work, the control allocation problem is formulated based on the reduced-order model and it is assumed that this model accurately represents the dominant contribution of different actuators to the inter-area modes of interest.\\ \indent
The Hankel norm approximation \cite{KGLOVER} can be used to obtain the reduced-order model and the order of model reduction can be determined by examining the Hankel singular values. Considering the reduced-order system with state variables $x_r\in R^{n}$, using an appropriate transformation $z=\psi x_r$ where $\psi \in R^{n \times n}$, the realization in modal form can be written as 
\begin{IEEEeqnarray}{rCl}
	\dot{z}(t) &=& \Lambda z(t)+\psi B_r u(t)  \\
	y(t) &=& C_r \psi^{-1} z(t) \\
	\Lambda &=& \begin{bmatrix} 
		\iota_1 & 0  & 0  & \dots \\
		0 &  \sigma_1 & \omega_1 \\
		0 & -\omega_1 & \sigma_1  \\
		\vdots	& & & \ddots
	\end{bmatrix}
\end{IEEEeqnarray}
where $\Lambda=\psi A_r \psi^{-1}$ is a block diagonal matrix whose elements are eigenvalues of $A_r$ (assuming no repeated eigenvalues), $u\in R^{m}$ denotes the input and $y\in R^{p}$ is the measured output. Real eigenvalue $\iota_i$ appears on diagonal and complex conjugate eigenvalues $\sigma_i \pm \omega_ij$ appear as a 2-by-2 block on the diagonal of $\Lambda$. By introducing the virtual control input $v\in R^{n}$, the system equations can be expressed as
\begin{IEEEeqnarray}{rCl}
	\dot{z}(t) &=& \Lambda z(t)+I_n v(t)  \label{eq:1} \\
	y(t) &=& C_r \psi^{-1} z(t) 		  \label{eq:2}\\
	v(t) &=& \psi B_r u(t)    	 		  \label{eq:3}
\end{IEEEeqnarray}
which decomposes the system into two parts and leads to a modular design where WADC generates the virtual control signal $v$ and control allocator distributes the effort among the available actuators. Matrix $\psi$ is full rank and $rank (\psi B_r) =n < m$, hence $\psi B_r$ has null space of dimension $m-n$ in which $u$ can be perturbed without affecting the response
\subsection{Wide-area Damping Controller Design}
We designed a multi-objective damping controller based on LMI optimization technique introduced in \cite{CScherer} but our approach to the CA can accommodate other control approaches. The controller is designed based on the reduced order model (\ref{eq:1}) and (\ref{eq:2}) to avoid feasibility problems and realize practical low-order controllers. Further details of this approach to design WADC can be found in \cite{MERaoufat_1}. The damping controller designed by the above methodology can be written as:
\begin{IEEEeqnarray}{rCl}
	\dot{x}_k(t) &=& A_k{x}_k(t)+B_k y(t)\\
	v(t) &=& C_k{x}_k(t)+D_k y(t)
\end{IEEEeqnarray}
Although the above WADC is designed using robust control methods, failure in the communication links or in the actuators will lead to poor damping performance.
\subsection{Modal-based Sparse Control Allocation}
Based on the order of the reduced model, the system can now represent an over-actuated system and the problem of modal-based sparse control allocation with proper filtering to reduce the variations can be represented as follows
\begin{IEEEeqnarray}{rCl}
	\begin{aligned}	
		& \underset{u_t}{\text{min}}
		& & \lVert W_u u_t \lVert^2_2 + \lVert W_s \big(u_t-u_{t-T_s}\big) \lVert^2_2 + \lambda \lVert u_t \lVert_1 \\
		& \text{s. t.}  & &  \psi B_r u_t=v_t\\
		&				& &	 u_{\min} \leq u_t \leq u_{\max}
	\end{aligned}
	\label{eq:prob1}
\end{IEEEeqnarray}
where $W_u$ and $W_s$ are positive definite matrices, usually diagonal, and represent the weighting for distributions and variations in the control signal, respectively. The term $\lVert u_t \lVert_1= \sum_{i=1}^{m} \vert u_{t,i} \vert$ denotes the $\ell_1$ norm of control vector $u_t$ and $\lambda \geq 0$ is the regularization parameter. Virtual control input $v_t$ is derived from the nominal WADC at time $t$ and $T_s$ denotes the time step. The key feature of the proposed control allocation strategy is that the temporal sparse control vector $u_t$ is directed to actuators considering total cost, actuator rates, modal effects and actuator limitations, which leads to a constrained optimization problem (\ref{eq:prob1}). This method is based on prior knowledge of control limits and CA only communicates with necessary actuators to achieve the damping requirement. \\ \indent
The cost function of the above optimization can then be simplified to 
\begin{IEEEeqnarray}{l}
	\lVert W_u u_t \lVert^2_2 + \lVert W_s (u_t-u_{t-T_s}) \lVert^2_2 + \lambda \lVert u_t \lVert_1 \\ \nonumber
	\quad = u_t^T W^2_u u_t + (u_t-u_{t-T_s})^T W^2_s (u_t-u_{t-T_s})  + \lambda \lVert u_t \lVert_1			 \\ \nonumber
	\quad = u^T_t(W^2_u+W^2_s)u_t - 2u^T_t  W^2_s u_{t-T_s} + \lambda \lVert u_t \lVert_1 + \mathrm{const.} 	 \\ \nonumber	
	\quad = \lVert W(u_t-u_d) \lVert^2_2 + \lambda \lVert u_t \lVert_1 + \mathrm{const.} 
\end{IEEEeqnarray}
where	
\begin{IEEEeqnarray}{l}
	u_d=W^2_s (W^2_u + W^2_s)^{-1} u_{t-T_s} , \quad
	W=(W_u^2+W_s^2)^\frac{1}{2} \quad
	\label{eq:u_d,W}
\end{IEEEeqnarray}
Since constant terms in the objective function will not affect the optimal solution, they can be removed and the optimization can be cast in the form of least square optimization with $\ell_1$ norm regularization
\begin{IEEEeqnarray}{rCl}
	\begin{aligned}	
		& \underset{u_t}{\text{min}}
		& & \lVert W \big(u_t-u_d\big) \lVert^2_2  + \lambda \lVert u_t \lVert_1 \\
		& \text{s. t.}  & &  \psi B_r u_t=v_t\\
		&				& &	 u_{\min} \leq u_t \leq u_{\max}
	\end{aligned}
\end{IEEEeqnarray}
with $u_d$ and $W$ from (\ref{eq:u_d,W}). The problem can be approximated by utilizing the first constraint in the cost function using the Lagrangian multiplier $\rho$ and weighting function $W_v$.
\begin{IEEEeqnarray}{rCl}
	 \lVert W \big( u_t-u_d\big) \lVert^2_2 + \rho^2  \lVert W_v \big( \psi B_r u_t-v_t\big) \lVert^2_2 + \lambda \lVert u_t \lVert_1 = \qquad \nonumber \\ 
	\bigg \lVert   \begin{bmatrix} 	\rho W_v \psi B_r   \\ W  \end{bmatrix} u_t -
	    \begin{bmatrix}  \rho W_v v_t \\ W u_d     \end{bmatrix} \bigg \lVert^2_2 +  \lambda \lVert u_t \lVert_1	\quad \quad
\end{IEEEeqnarray}
Finally, we obtain the following optimization problem
\begin{IEEEeqnarray}{rCl}
	\begin{aligned}	
		& \underset{u_t}{\text{min}}
		& & \bigg \lVert   \begin{bmatrix} 	\rho W_v \psi B_r   \\ W  \end{bmatrix} u_t -
		\begin{bmatrix}  \rho W_v v_t \\ W u_d     \end{bmatrix} \bigg \lVert^2_2 +  \lambda \lVert u_t \lVert_1	\quad \\
		& \text{s. t.}  & &	 u_{\min} \leq u_t \leq u_{\max}
	\end{aligned}
\end{IEEEeqnarray}
\indent  In the control literature, there exist other methods to distribute the control signal based on cost \cite{OHärkegard_1} or actuator limits \cite{JBDavidson}, but these have not considered the effects on modal system or sparsity. This technique allows us to give the highest priority to the control efforts associated with the critical inter-area modes by using the weighting function $W_v$ and obtain the feasibility regions in modal coordinates. By decomposing the control vector $u_t$ to positive and negative components, we introduce nonnegative variables $u^+_t$, $u^-_t$ and $q_t=\begin{bmatrix}	u^+_t & u^-_t \end{bmatrix}^T$ such that
\begin{IEEEeqnarray}{rCl}
	u_t = u^+_t - u^-_t = \begin{bmatrix} I_m & -I_m \end{bmatrix} q_t ; \qquad		 u^+_t, u^-_t \geq 0 \quad
\end{IEEEeqnarray}
The $\ell_1$ norm can then be modeled as $ \lVert u_t \lVert_1 = \bar{1}^T q_t$ (with $\bar{1}$ being a vector of ones) and the $\ell_1$-regularized least square problem can be transformed into a quadratic programing with simple box constraints as follow
\begin{IEEEeqnarray}{rCl}
	\begin{aligned}	
		& \underset{q_t}{\text{min}}
		& & q_t^T \begin{bmatrix} \mathcal{A}^T \mathcal{A} & -\mathcal{A}^T \mathcal{A} \\ -\mathcal{A}^T \mathcal{A} & \mathcal{A}^T \mathcal{A} \end{bmatrix} q_t + ( 2\begin{bmatrix} -\mathcal{A}^T \mathcal{B} \\ \mathcal{A}^T \mathcal{B}\end{bmatrix}^T \! \! + \lambda \bar{1}^T ) q_t   \quad\\
		& \text{s. t.}  & &	 \bar{0} \leq q_t \leq \begin{bmatrix} u_{\max} \\ -u_{\min} \end{bmatrix} 
	\end{aligned}
	\label{eq:opt_2}
\end{IEEEeqnarray}
where	
\begin{IEEEeqnarray}{l}
	\mathcal{A}= \begin{bmatrix} 	\rho W_v \psi^T B_r   \\ W  \end{bmatrix}, \quad
	\mathcal{B}= \begin{bmatrix}  \rho W_v v_t \\ W u_d     \end{bmatrix} 
\end{IEEEeqnarray}
For most problems, these quadratic programs can be solved efficiently using interior-point or active-set methods. Note the transformed problem is an optimization over $2m$-dimensional vector space.
\section{Dynamic Model of the WECC Test System}
\begin{figure}[t]
	\centering \vspace{-0.2cm}
	\includegraphics[width=3.5in]{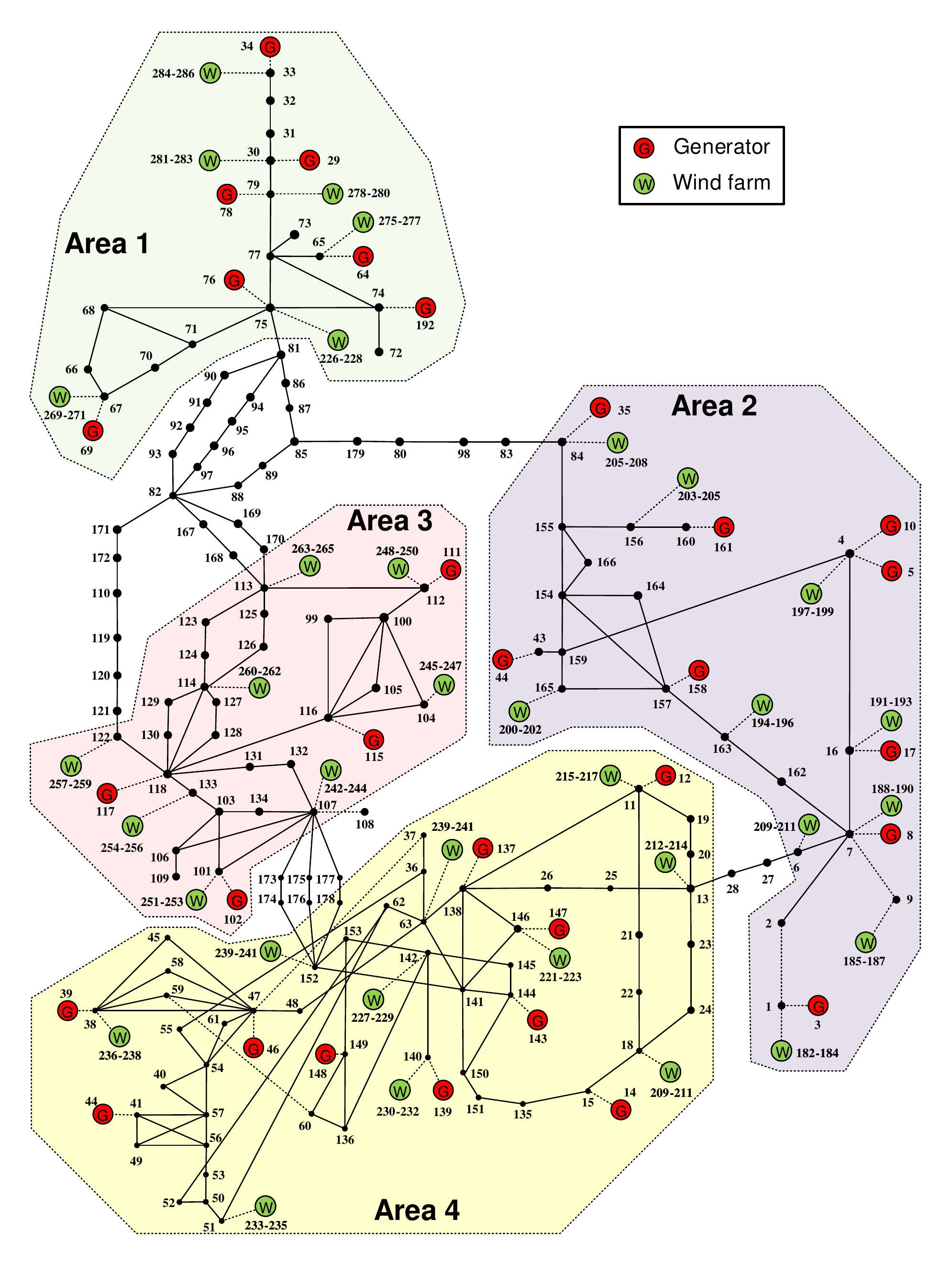} \vspace{-0.5cm}
	\caption{One-line diagram of a WECC power system with Type 4 wind farms.}
	\label{WECC_Diagram} \vspace{-0.4cm}
\end{figure}

\begin{figure}[t]
	\centering
	\includegraphics[width=3.5in]{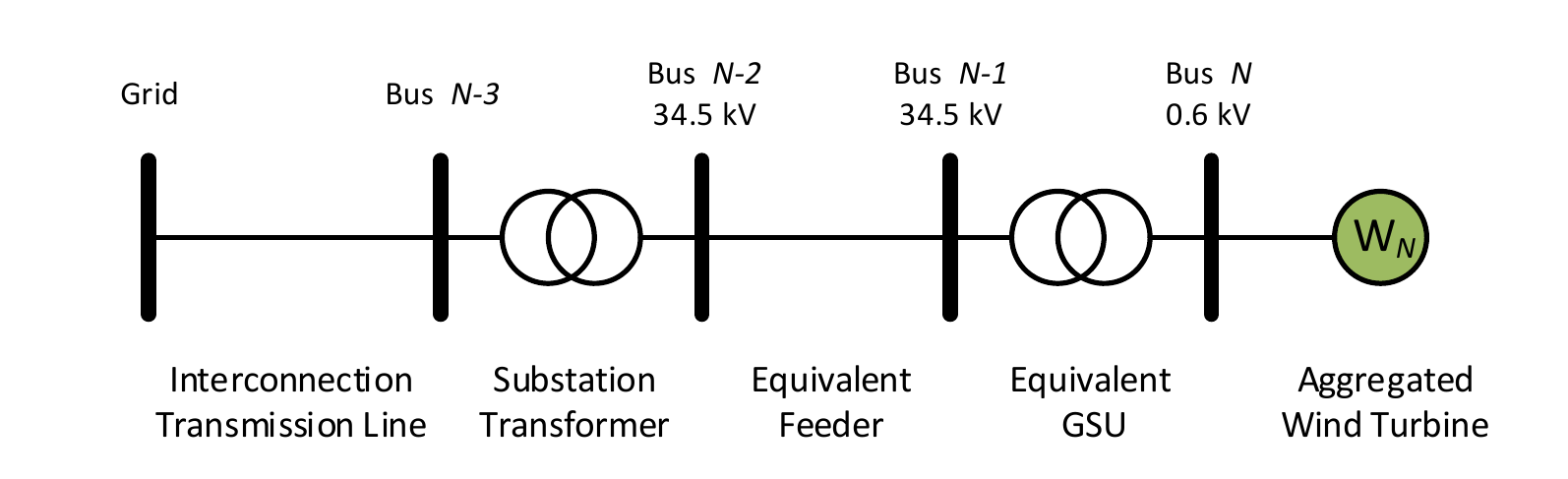} \vspace{-0.5cm}
	\caption{Single-generator representation of aggregated wind farm model.}
	\label{Wind_Circuit} \vspace{-0.3cm}
\end{figure}

A modified $286$-bus WECC system is used in this study to capture the effects of redundant actuators over the inter-area modes. As shown in Fig. \ref{WECC_Diagram}, this system consists of $31$ synchronous generators with generation of $60.25$ GW and $35$ small-scale wind farms, each rated at $60$ MVA and $50$ MW, with total generation of $1.75$ GW. Each generator is represented using a two-axis model equipped with a high-gain AVR system and a power system stabilizer (PSS1A) to damp the local oscillation modes \cite{JMachowski}. All loads are assumed to be constant power and original parameters regarding the network data and operating conditions are given in \cite{WECC}. \\ \indent
Wind farms are represented by an aggregated model of Type $4$ wind turbines. In this work, the base power of each wind farm is scaled based on the total number of wind turbines while the parameters are assumed to be constant. The equivalent circuit is shown in Fig. \ref{Wind_Circuit} and further details on network and model parameters can be found in \cite{WECC_Wind}. In this study, the damping controller is performed by adding a supplementary signal $u$ to the reactive power control loop for reactive power modulation. We assume each wind farm is constrained to run within a specific power factor range, for example $0.9$ lagging to $0.9$ leading which is typical for Type $4$ machines \cite{WChristiansen}. As a result, a hard limits of $u_{\max}=-u_{\min}=0.4$ pu are imposed on the supplementary signal of each wind farm.
\section{Numerical Results}
Detailed studies based on a nonlinear model of the WECC system described in the previous section are performed to verify the performance of the proposed control allocation method.
\subsection{Linear Analysis and Design of WADC}
\begin{figure}[t]
	\centering
	\includegraphics[width=3.4in]{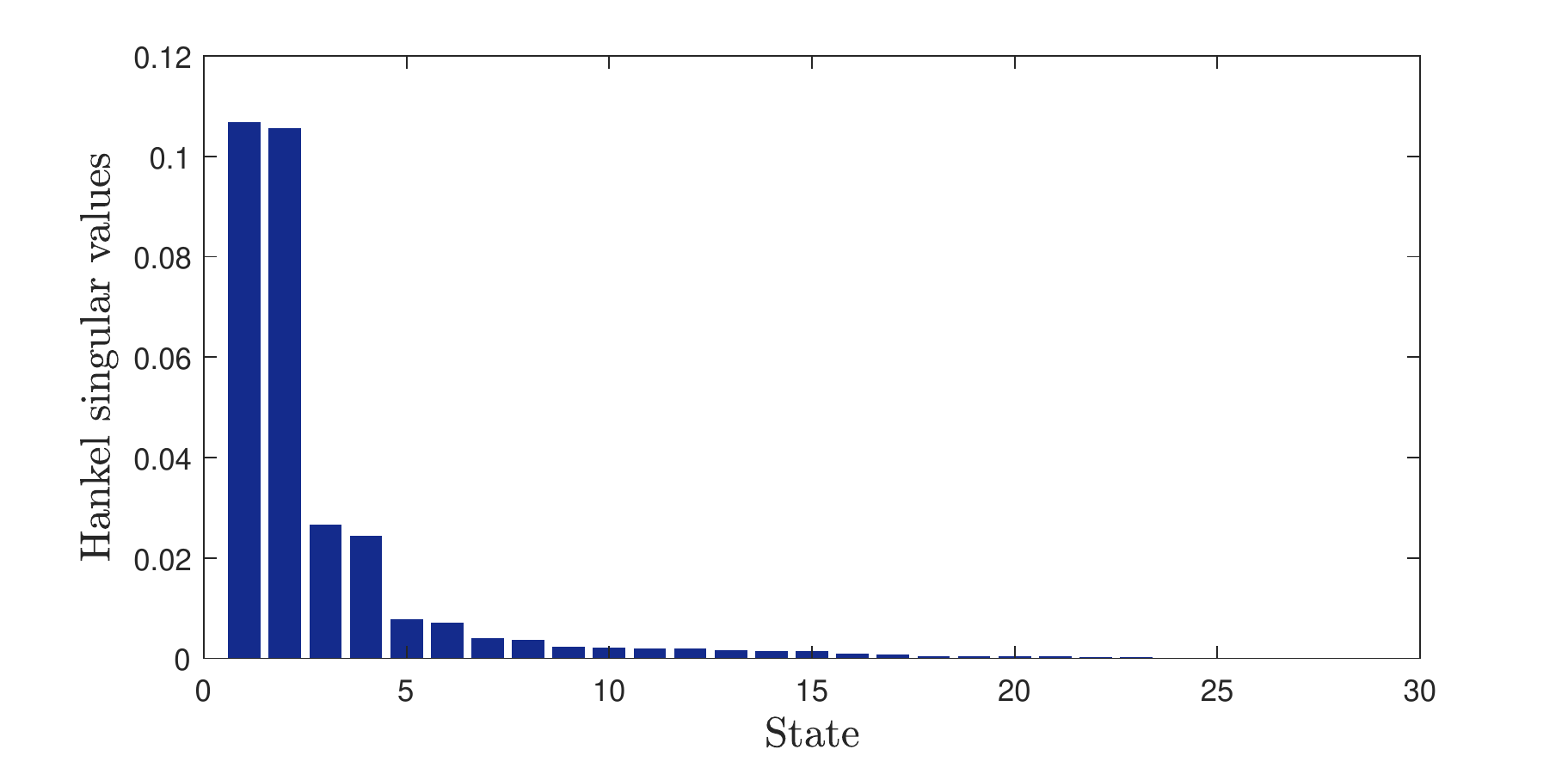} \vspace{-0.2cm}
	\caption{Hankel singular values of the modified WECC system.}
	\label{Hankel_SV} \vspace{-0.2cm}
\end{figure}

This system exhibits several low-frequency oscillation modes that are characterized in Table \ref{Table:WECC_Mode}. Critical mode $3$ with frequency of $0.564$ Hz and a low damping ratio of $0.98$\% is of high interest and represents the inter-area mode between area $2$ and $4$. Based on an observability measure, speed deviation of $G_{10}$ is selected as the best candidate signal for our controller as it has the highest observability over the critical mode (details of this approach are given in \cite{MERaoufat_1}). The test system has $490$ states and the order of the reduced model is chosen as $n=6$ to preserve the largest Hankel singular values \cite{KGLOVER} as shown in Fig \ref{Hankel_SV}. The WADC is designed based on the $6^{th}$-order model to meet or exceed $6$\% damping over the inter-area modes.
\begin{table}[!t]
	\renewcommand{\arraystretch}{1.2}
	\caption{Low-Frequency Modes of the Modified WECC System, Base Case with no Controllers.}
	\centering
	\begin{tabular}{|c|c|c|c|c|c|}
		\hline
		Mode No. & Participating Generators & Freq. (Hz)    &  Damp. ($\%$)    \\ \hline
		1 & Area 1 vs. Area 2,4      & 0.327  &  6.99    \\ \hline 
		2 & $G_{34}$ vs. $G_{64}$    & 0.442  &  11.62   \\ \hline
		3 &  Area 2 vs. Area 4       & 0.564  &  0.98    \\ \hline
	\end{tabular}
	\label{Table:WECC_Mode}
\end{table}
\subsection{Design of Modal-Based Control Allocation}
The proposed CA is implemented as a user-defined model (UDM) in TSAT \cite{TSAT} and the optimization algorithm (\ref{eq:opt_2}) is performed using dynamically linked blocks (DLBs) and MATLAB with a fixed time step of $T_s=0.02$ s and interior-point method. The available small-scale wind farms are chosen as the set of redundant actuators as follows
\begin{IEEEeqnarray}{rCl}
	\mathcal{R}=\big\{W_{184}, W_{187}, W_{190}, \dots,  W_{286} \big\}
\end{IEEEeqnarray}
where $i^{th}$ element of vector $\mathcal{R}$ is associated with the $i^{th}$ column of matrix $B_r$. The weighting functions and gains are chosen as $W_u:=I_{35}$, $W_s:=2 W_u$, $W_v:=\textrm{diag}(2,2,4,4,8,8)$, $\lambda:=1$ and $\rho:=100$. This choice of weighting matrix $W_v$ gives the highest priority to the control efforts regarding the critical mode $3$. Moreover, the weighting $W_u$ can also be chosen based on the reliability of each actuator and the corresponding communication link. 
\subsection{Nonlinear Simulations}
\begin{figure}[t]
	\centering \vspace{-0.2cm} 
	\includegraphics[width=3.55in]{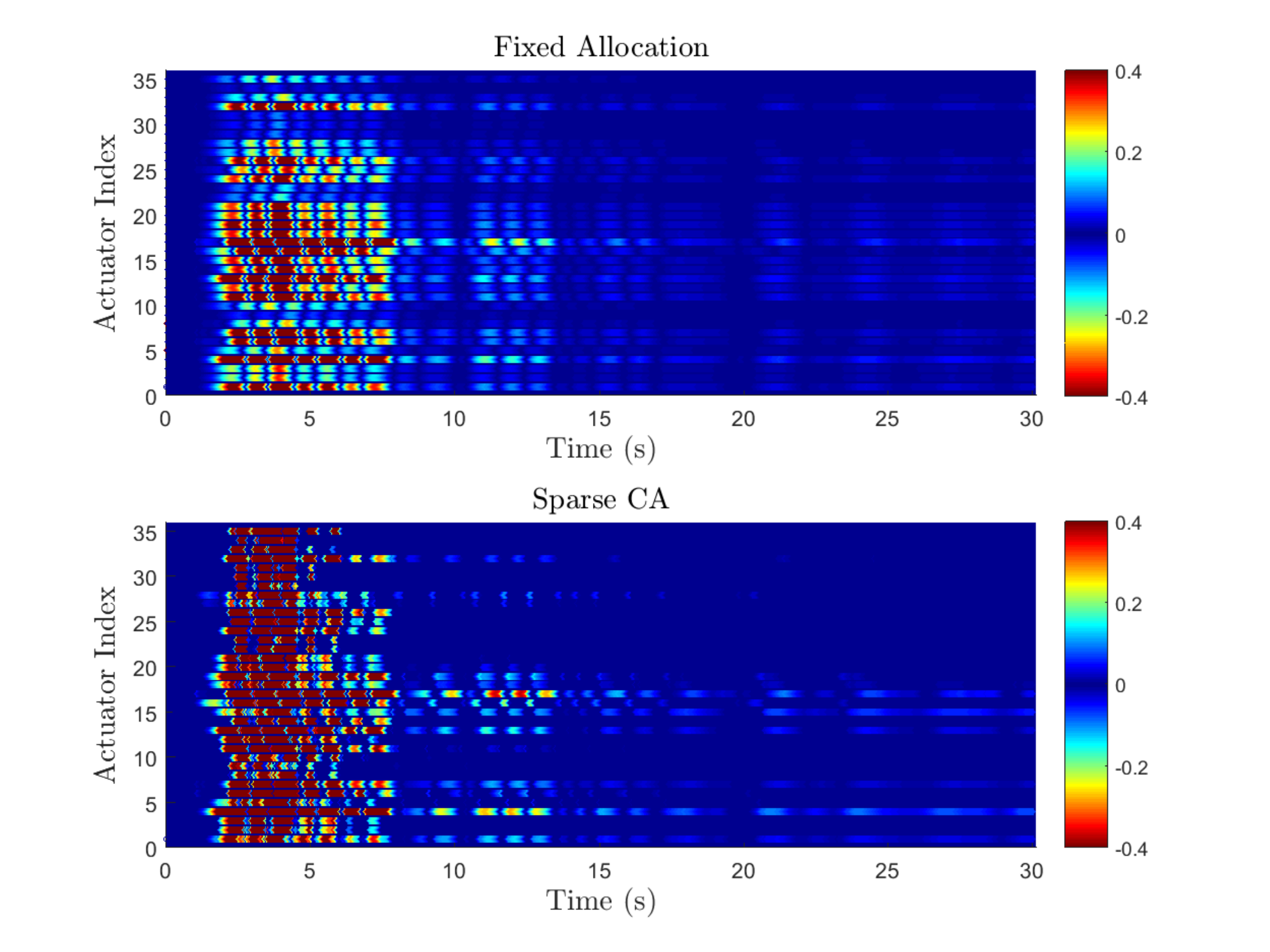} \vspace{-0.7cm}  
	\caption{Supplementary control signal $u$ for different actuators in response to three-phase fault at bus \#$139$ in case A.}
	\label{Control_Case_A} \vspace{-0.4cm}  
\end{figure}

\begin{figure}[t] \vspace{-0.7cm}  
	\includegraphics[width=3.5in]{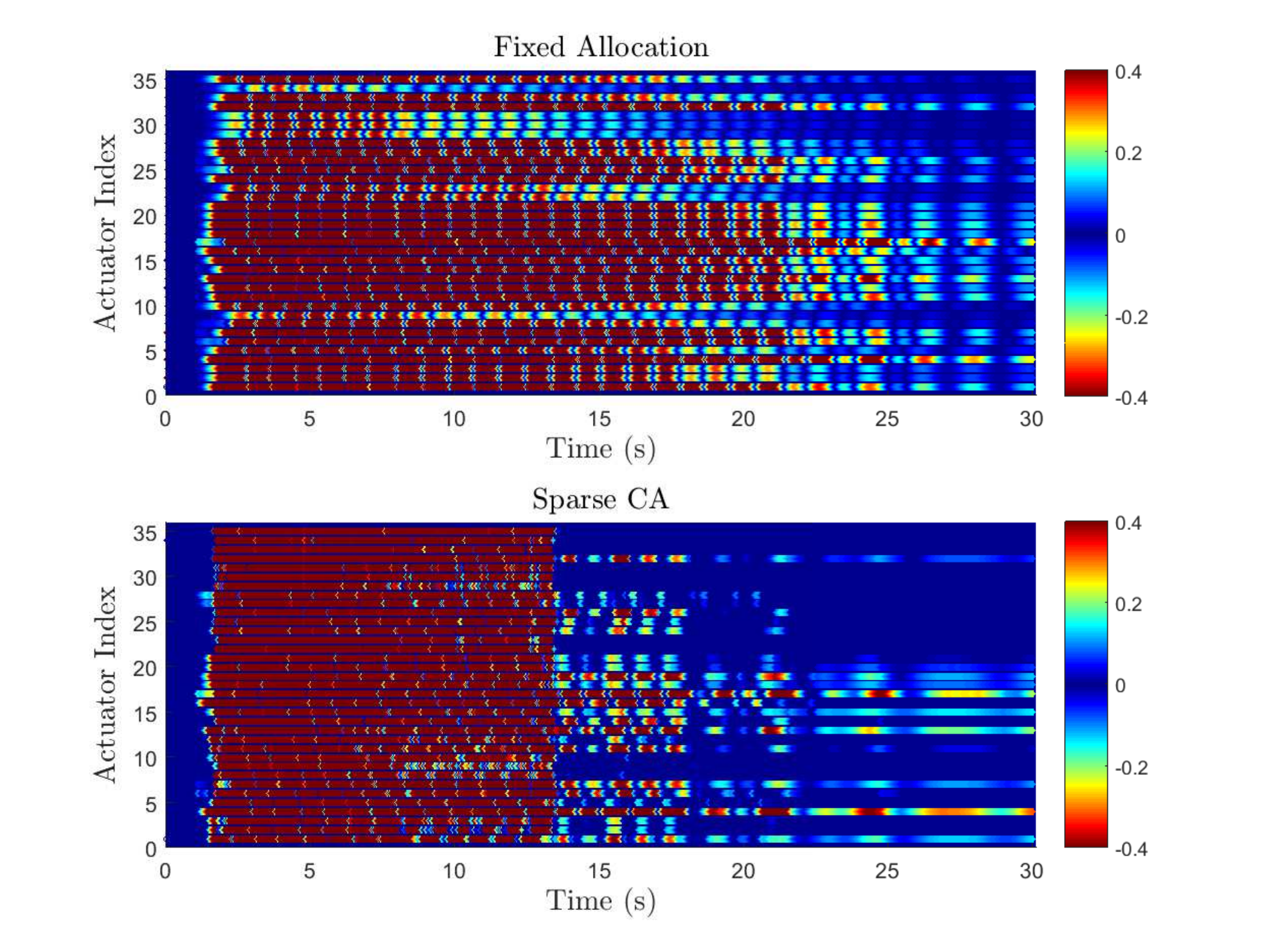} \vspace{-0.7cm}  
	\caption{Supplementary control signal $u$ for different actuators in response to three-phase fault at bus \#$139$ in case B.}
	\label{Control_Case_B} \vspace{-0.4cm}  
\end{figure}

Nonlinear transient studies were performed using TSAT and Prony analysis is used to extract the damping coefficient of the inter-area oscillation based on the nonlinear response. In this study, the time frame of analysis (oscillation) is restricted to a few seconds, so it is reasonable to assume that the wind speed remains effectively constant during this period. Cases of interest include faults in both the physical system and actuators. In the physical system, a symmetrical three-phase fault is applied at bus \#$139$, which is a severe disturbance, to stimulate the critical inter-area mode.  \\ \indent
\begin{figure}[t]
	\centering \vspace{-0.2cm}
	\includegraphics[width=3.43in]{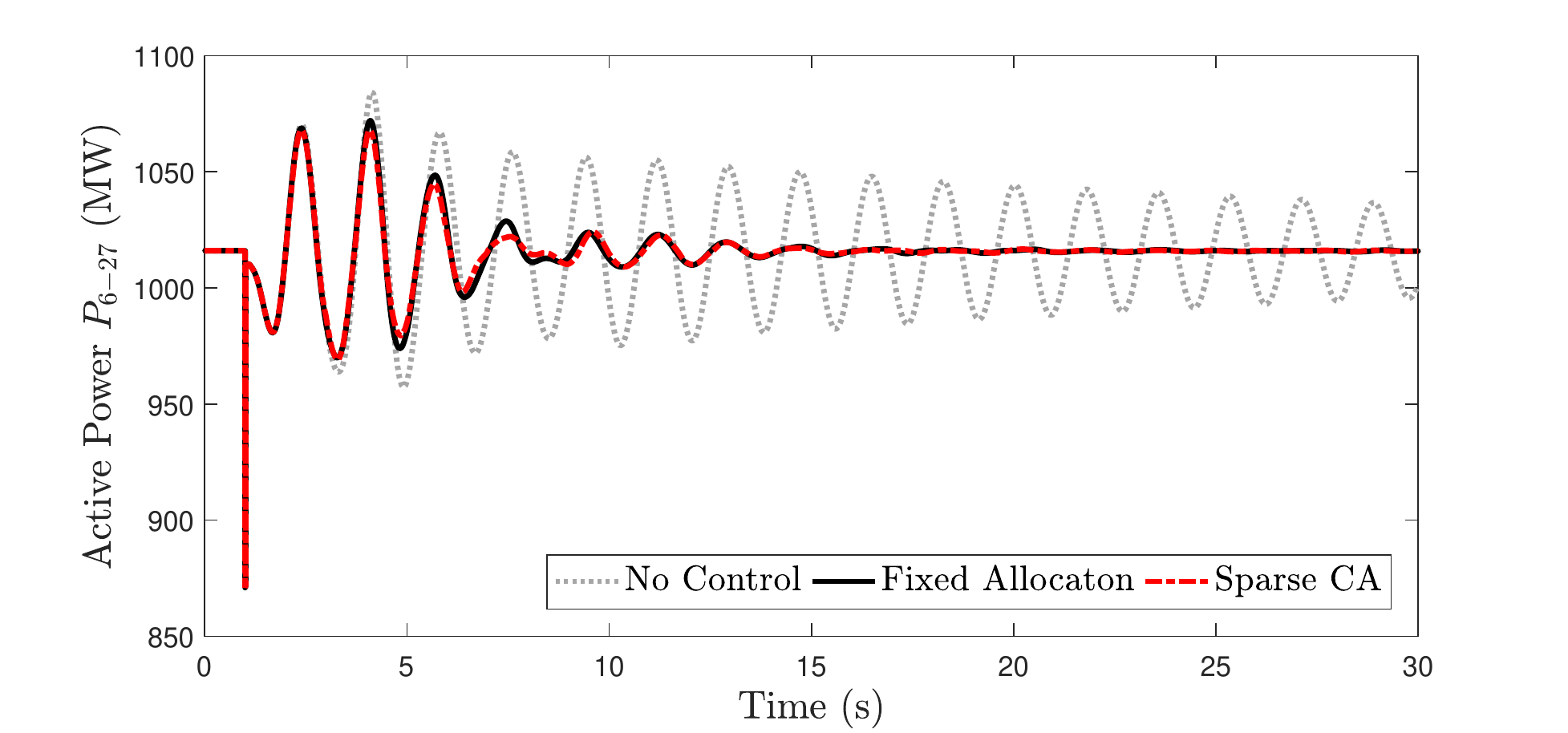}   \vspace{-0.1cm}     
	\caption{Responses of the WECC system to three-phase fault at bus \#$139$ in case A.}
	\label{Power_Case_A}
	
	\centering
	\includegraphics[width=3.43in]{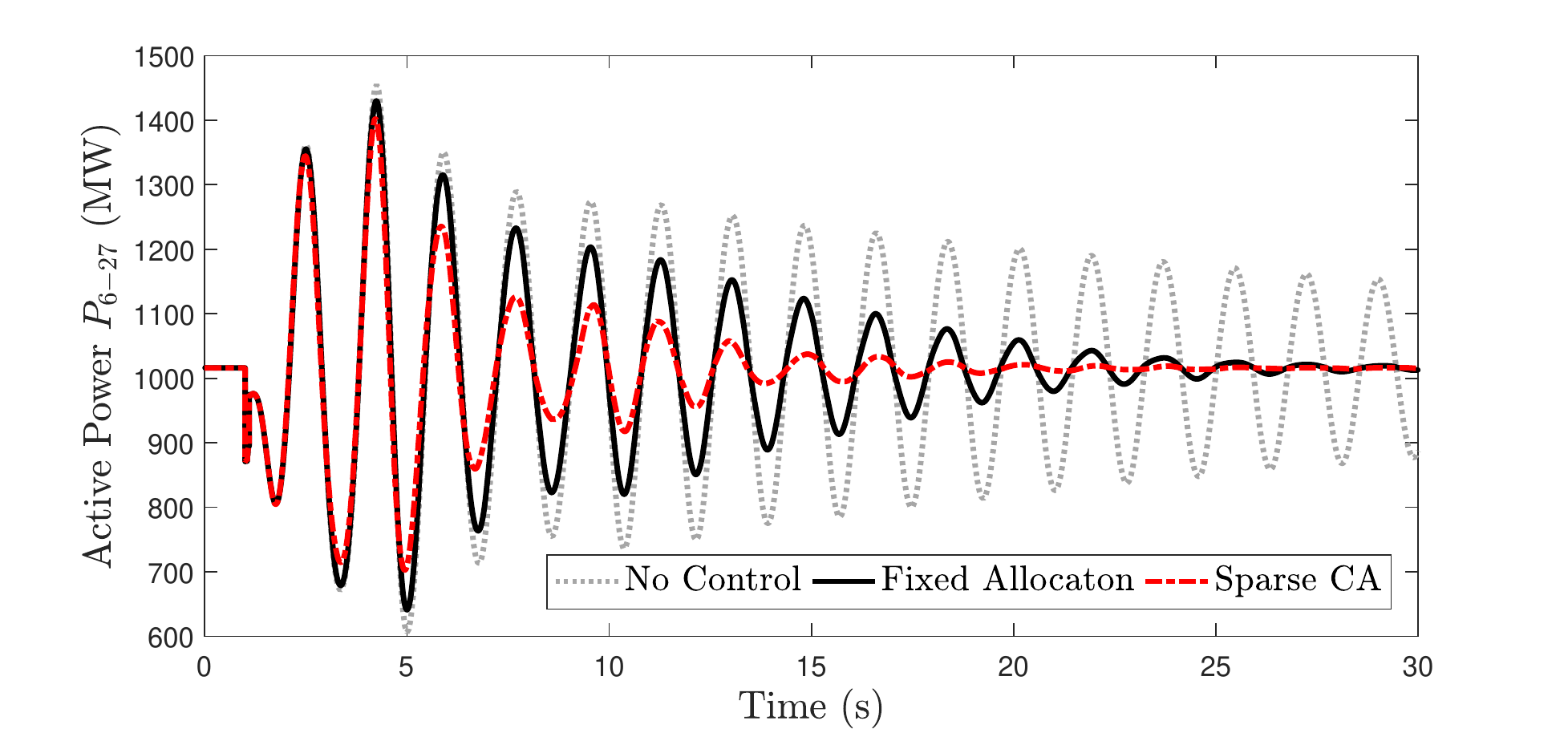}   \vspace{-0.1cm}    
	\caption{Responses of the WECC system to three-phase fault at bus \#$139$ in case B.}
	\label{Power_Case_B}

	\centering	
	\includegraphics[width=3.43in]{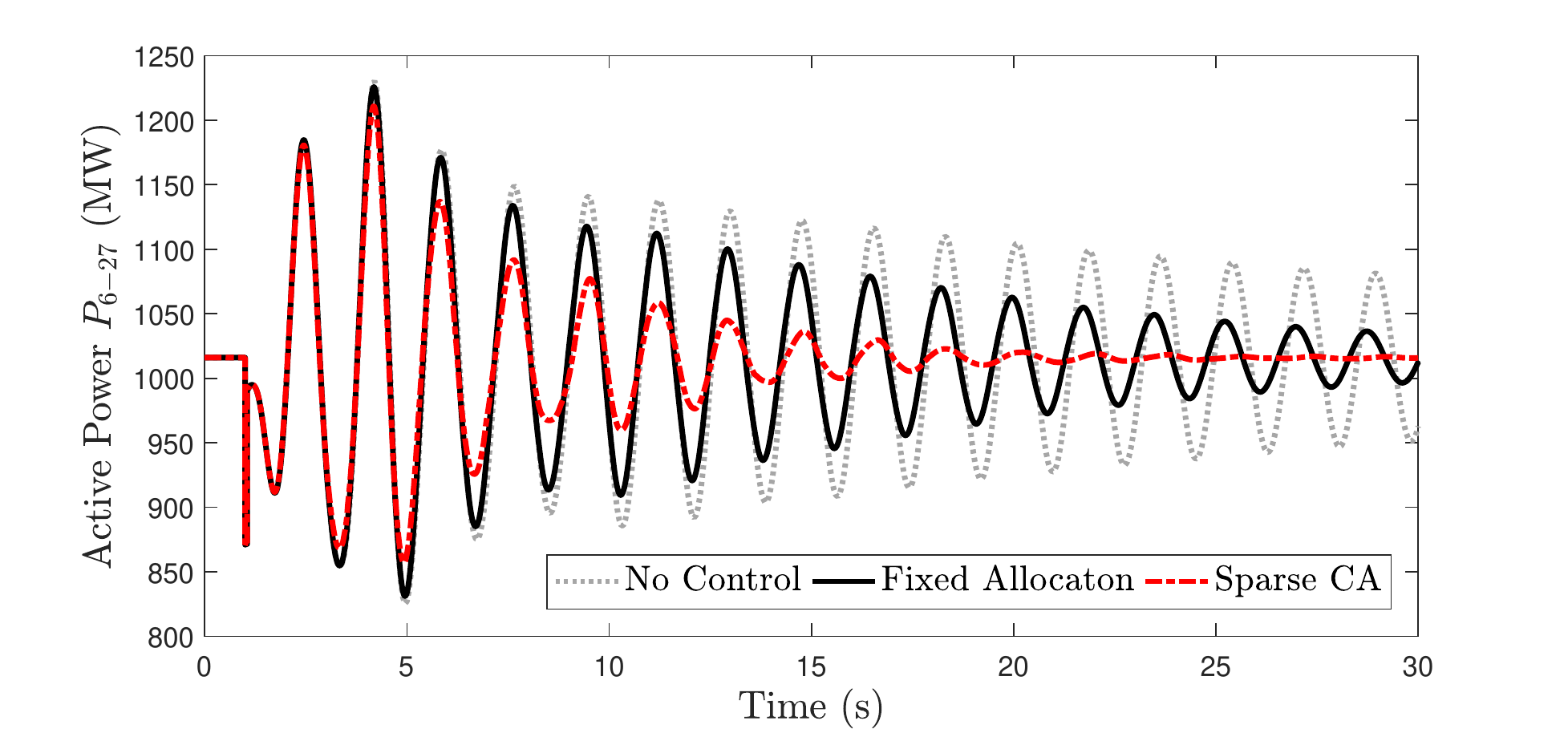}	  \vspace{-0.1cm}      
	\caption{Responses of the WECC system to three-phase fault at bus \#$139$ in case C.}
	\label{Power_Case_C} \vspace{-0.3cm}   
\end{figure}
To illustrate the benefit of sparse control allocation, three control cases were evaluated and compared during transient response. First, the system with no control is considered. Second, a WADC with fixed allocation $ u(t)=(\psi B_r)^{\dagger} v(t) $ is considered based on pseudo-inverse calculation. Finally, a sparse control allocation is considered to include hard limits and actuator status in the design. Active power of the inter-area transmission line $6-27$ is shown in Figs \ref{Power_Case_A}, \ref{Power_Case_B} and \ref{Power_Case_C} for the following cases
\begin{itemize}
	\item Case A: No faulty actuators and fault duration of $1$ cycle;
	\item Case B: No faulty actuators and fault duration of $6$ cycles;
	\item Case C: Faults in $70$\% of the actuators and fault duration of $3$ cycles;
\end{itemize}
It can be seen that in case A, where fault duration is short and the required control effort is less, both sparse CA and fixed allocation method can improve the damping to $7.2$\% compared to the open-loop damping of $0.98$\%. In case B, where fault duration is longer and requires extensive control efforts, the damping ratio of the fixed allocation method reduces to $2.82$\%. However, the sparse CA achieves a damping ratio of $5.55$\% as it considers the actuator limitations in control redistribution. In case C with a shorter fault duration but $70$\% actuator failures (either from multiple wind farms are disconnected, communication congestions, changes in wind speed), the sparse CA will again dampen the oscillations by redistributing the control signal to healthy actuators and maintain sufficient damping of $4.89$\% compared to $2.03$\% under a fixed allocation. Comparing these results, it can be seen that the proposed method enhances fault-tolerance of the WADC system.  \\ \indent
Figs. \ref{Control_Case_A} and \ref{Control_Case_B} illustrate the sparse CA outputs in case A and B, respectively. It can be seen that the control signal $u$ is temporally sparse relative to the fixed control allocation method. Additional results for different actuator fault combinations are presented in Table \ref{table:Faults}. In all cases, the physical fault is assumed to be with a duration of $3$ cycles. It can be observed that our proposed method tolerates various combinations of failures and maintains a higher damping ratio over the critical inter-area mode.
\begin{table}[t] \vspace{-0.4cm}  
	\renewcommand{\arraystretch}{1.1}
	\caption{Damping Ratio of the Critical Mode for Different Actuator Fault Combinations.}
	\centering
	\begin{tabular}{|c|c|c|c|}
		\hline
		\multicolumn{1}{|c|}{Actuator} & \multicolumn{3}{c|}{Damping Ratio ($\%$)}          \\ \cline{2-4} 
		\multicolumn{1}{|c|}{Failure ($\%$)}  & \multicolumn{1}{c|}{Sparse CA} & \multicolumn{1}{c|}{Fixed Alloc.}  & \multicolumn{1}{c|}{No Control}\\ \hline
		 0  &   6.79  &  5.12  &  0.97      \\ \hline	
		10  &   6.56  &  4.62  &  0.97      \\ \hline	
		30  &   6.43  &  4.01  &  0.97      \\ \hline   
		50	&   5.80  &  3.04  &  0.97      \\ \hline   
		70	&   4.89  &  2.03  &  0.97      \\ \hline   
		80	&   4.33  &  1.59  &  0.97      \\ \hline   
	\end{tabular}
	\label{table:Faults} 
\end{table}
\section{Conclusions}
This work proposes a sparse control allocation technique for fault-tolerant wide-area damping controllers and coordinated control of multiple actuators. This method leads to a modular design process where the damping controller generates the virtual control signal and the supervisory CA distributes the control efforts to the necessary actuators based on the desired control actions, actuator limits and modal effects. The proposed approach is applied to a modified $286$-bus Western Electricity Coordinating Council (WECC) system with distributed small-scale wind farms. Simulation results show significant improvement in resiliency due to various system failures.
\bibliography{IEEEabrv_Ehsan,Ref}  
\bibliographystyle{IEEEtran}
\end{document}